\documentclass[letterpaper,12pt,notoc]{JHEP3}
\def\pd{\partial}
\def\mc{\mathcal}

\usepackage{graphicx}
\usepackage{amsmath}
\usepackage{amssymb}
\preprint{ \hbox{}\hfill arXiv: 1507.01515}

\title{Twisted compactification of $N=2$ 5D SCFTs to three
and two dimensions from $F(4)$ gauged supergravity}
\author{Parinya Karndumri\\
String Theory and Supergravity Group, Department
of Physics, Faculty of Science, Chulalongkorn University, 254 Phayathai Road, Pathumwan, Bangkok 10330, Thailand\\
E-mail: \email{parinya.ka@hotmail.com}}

\abstract{We study supersymmetric $AdS_4\times \Sigma_2$ and
$AdS_3\times \Sigma_3$ solutions in half-maximal gauged supergravity
in six dimensions with $SU(2)_R\times SU(2)$ gauge group. The gauged
supergravity is obtained by coupling three vector multiplets to the
pure $F(4)$ gauged supergravity. The $SU(2)_R$ R-symmetry together
with the $SO(3)\sim SU(2)$ symmetry of the vector multiplets are
gauged. The resulting gauged supergravity admits supersymmetric
$AdS_6$ critical points with $SO(4)\sim SU(2)\times SU(2)$ and
$SO(3)\sim SU(2)_{\textrm{diag}}$ symmetries. The former corresponds
to five-dimensional $N=2$ superconformal field theories (SCFTs) with
$E_1\sim SU(2)$ symmetry. We find new classes of supersymmetric
$AdS_4\times \Sigma_2$ and $AdS_3\times \Sigma_3$ solutions with
$\Sigma_{2,3}$ being $S^{2,3}$ and $H^{2,3}$. These solutions
describe SCFTs in three and two dimensions obtained from twisted
compactifications of the aforementioned five-dimensional SCFTs with
different numbers of unbroken supersymmetry and various types of
global symmetries.}


\begin{document}
\section{Introduction}
Field theories in six and five dimensions have been shown to posses
non-trivial conformal fixed points
\cite{Seiberg_5Dfield,Seiberg_6D_fixed_points}. However, higher
dimensional superconformal field theories (SCFTs) are not well
understood as their lower dimensional analogues. The study of
five-dimensional SCFTs using the AdS/CFT correspondence
\cite{maldacena} has attracted a lot of attention both from ten and
six-dimensional point of views, see for example
\cite{D4D8,Bergman,Bergman2,F4_nunezAdS6,F4_flow,5DSYM_from_F4}. And
recently, the investigation of supersymmetric $AdS_6$ solutions has
been carried out systematically in
\cite{AdS6_from10D,AdS6_Tomasiello,AdS_6_Kim}.
\\
\indent An approach to understand higher dimensional field theories
is to make some compactification of these theories to lower
dimensions. The resulting lower dimensional field theories
preserving some supersymmetry are usually obtained by twisted
compactifications, and the holographic study via the AdS/CFT
correspondence is still applicable at least in the large $N$ limit
\cite{MN_nogo}. From string/M theory point of view, these twisted
field theories can be interpreted as wrapped branes on certain
curved manifolds. In many cases, there is a description in terms of
lower dimensional gauged supergravities. In particular, for the
present case of five-dimensional SCFTs, the effective supergravity
theory is the $N=(1,1)$ $F(4)$ gauged supergravity and its
matter-coupled version \cite{ferrara_AdS6}.
\\
\indent In this work, we will explore some aspects of twisted
compactifications of five-dimensional SCFTs within the framework of
half-maximal gauged supergravity in six dimensions coupled to matter
multiplets \cite{F4SUGRA1,F4SUGRA2}. A similar study in the pure
$F(4)$ gauged supergravity \cite{F4_Romans} have been carried out in
\cite{F4_nunez} in which some $AdS_4\times \Sigma_2$ and
$AdS_3\times \Sigma_3$ solutions have been identified along with
their possible dual field theories. We will further investigate
solutions of this type in the matter-coupled $F(4)$ gauged
supergravity. This could presumably give rise to more general
solutions than those given in \cite{F4_nunez}. The result would also
provide new solutions describing IR fixed points of the RG flows
from SCFTs in five dimensions to three and two-dimensional SCFTs
with different numbers of supersymmetry.
\\
\indent As a starting point, we add three vector multiplets to the
$F(4)$ gauged supergravity resulting in an $SU(2)_R\times SU(2)\sim
SO(3)_R\times SO(3)$ gauge group with the first factor being the
R-symmetry group. $AdS_6$ vacua of this theory including possible
holographic RG flows between the dual SCFTs and RG flows to
non-conformal field theories have already been studied in
\cite{F4_flow} and \cite{5DSYM_from_F4}. From the result in
\cite{F4_flow}, there are two supersymmetric $AdS_6$ critial points.
Both of them preserve the full sixteen supercharges, but one of
them, with non-vanishing scalar fields, break the full
$SU(2)_R\times SU(2)$ symmetry to its diagonal subgruop. These two
critical points are dual to certain $N=2$ SCFTs in five dimensions
by the usual AdS/CFT correspondence.
\\
\indent We then proceed by looking for possible $AdS_4\times
\Sigma_2$ and $AdS_3\times \Sigma_3$ solutions for $\Sigma_{2,3}$
being $S^{2,3}$ or $H^{2,3}$ with different residual symmetries. The
resulting solutions would be dual to SCFTs in three and two
dimensions obtained from twisted compactifications of the above
mentioned five-dimensional SCFTs. These will give new $AdS_4$ and
$AdS_3$ solutions from six-dimensional gauged supergravity and
provide appropriate gravity backgrounds in the holographic study of
gauge theories in five and lower dimensions.
\\
\indent The paper is organized as follow. We give a brief review of
the $F(4)$ gauged supergravity coupled to three vector multiplets in
section \ref{6D_SO4gaugedN2}. Possible supersymmetric $AdS_4$ and
$AdS_3$ solutions are given in section \ref{AdS4} and \ref{AdS3},
respectively. In section \ref{conclusion}, we give some conclusions
and comments about the results. We also include an appendix
describing supersymmetric $AdS_6$ critical points previously found
in \cite{F4_flow} as well as an analytic RG flow between them.
\section{Matter coupled $N=(1,1)$ $SU(2)\times SU(2)$ gauged supergravity in six dimensions}\label{6D_SO4gaugedN2}
In this paper, we are interested in $N=(1,1)$ gauged supergravity
with $SU(2)\times SU(2)$ gauge group. This gauged supergravity can
be obtained by coupling three vector multiplets to the pure $F(4)$
gauged supergravity constructed in \cite{F4_Romans}. The full
construction by using the superspace approach can be found in
\cite{F4SUGRA1, F4SUGRA2}. Apart from different metric signature
$(-+++++)$, we will mostly follow the notations and conventions
given in \cite{F4SUGRA1} and \cite{F4SUGRA2}.
\\
\indent The matter coupled $N=(1,1)$ gauged supersymmetry consists
of the supergravity multiplet given by
\begin{displaymath}
\left(e^a_\mu,\psi^A_\mu, A^\alpha_\mu, B_{\mu\nu}, \chi^A,
\sigma\right)
\end{displaymath}
and three vector multiplets with the field content
\begin{displaymath}
(A_\mu,\lambda_A,\phi^\alpha)^I.
\end{displaymath}
In the above expressions, $\psi^A_\mu$, $\chi^A$ and $\lambda_A$
denote the gravitini, the spin-$\frac{1}{2}$ fields and the
gauginos, respectively. All spinor fields $\chi^A$, $\psi^A_\mu$ and
$\lambda_A$ as well as the supersymmetry parameter $\epsilon^A$ are
eight-component pseudo-Majorana spinors with indices $A,B=1,2$
referring to the fundamental representation of the $SU(2)_R\sim
USp(2)_R$ R-symmetry. Space-time and tangent space indices are
denoted respectively by $\mu,\nu=0,\ldots ,5$ and $a,b=0,\ldots, 5$.
$e^a_\mu$ and $\sigma$ are the graviton and the dilaton.
$A^\alpha_\mu,\, \alpha=0,1,2,3$, are four vector fields in the
gravity multiplet. Three of these vector fields will be used to
gauge the $SU(2)_R$ R-symmetry. The index $I=1,2,3$ labels the three
vector multiplets, and finally $B_{\mu\nu}$ is the two-form field
which admits a mass term.
\\
\indent There are $13$ scalar fields parametrized by
$\mathbb{R}^+\times SO(4,3)/SO(4)\times SO(3)$ coset manifold in
which the $\mathbb{R}^+\sim SO(1,1)$ part corresponds to the
dilaton. Possible gauge groups are subgroups of the global symmetry
group $\mathbb{R}^+\times SO(4,3)$. In the present paper, we will
consider only the compact gauge group $SU(2)\times SU(2)\sim
SO(3)\times SO(3)$. The first factor is the $SU(2)_R$ R-symmetry
identified with the diagonal subgroup of $SU(2)\times SU(2)\sim
SO(4)\subset SO(4)\times SO(3)$. Following \cite{F4SUGRA1} and
\cite{F4SUGRA2}, we will decompose the $\alpha$ index into
$\alpha=(0,r)$ in which $r=1,2,3$. Indices $r,s$ will become adjoint
indices of the $SU(2)_R$ R-symmetry.
\\
\indent The $12$ vector multiplet scalars given by the
$SO(4,3)/SO(4)\times SO(3)$ coset can be parametrized by the coset
representative $L^\Lambda_{\phantom{as}\Sigma}$,
$\Lambda,\Sigma=0,\ldots , 6$. We can split the index $\Sigma$,
transforming by right multiplications of the local $SO(4)\times
SO(3)$ composite symmetry, in $L^\Lambda_{\phantom{as}\Sigma}$ to
$(L^\Lambda_{\phantom{as}\alpha},L^\Lambda_{\phantom{as}I})$ and
further to $(L^\Lambda_{\phantom{as}0}, L^\Lambda_{\phantom{as}r},
L^\Lambda_{\phantom{as}I})$. The vielbein of the
$SO(4,3)/SO(4)\times SO(3)$ coset $P^I_{\phantom{a}\alpha}$ and the
$SO(4)\times SO(3)$ composite connections
$\Omega^{\alpha\beta}=(\Omega^{rs},\Omega^{r0})$ can be obtained
from the left-invariant 1-form of $SO(4,3)$
\begin{equation}
\Omega^\Lambda_{\phantom{sa}\Sigma}=
(L^{-1})^\Lambda_{\phantom{sa}\Pi}\nabla L^\Pi_{\phantom{sa}\Sigma},
\qquad \nabla
L^\Lambda_{\phantom{sa}\Sigma}=dL^\Lambda_{\phantom{sa}\Sigma}
-f^{\phantom{s}\Lambda}_{\Gamma\phantom{sa}\Pi}A^\Gamma
L^\Pi_{\phantom{sa}\Sigma},
\end{equation}
with the following identification
\begin{equation}
P^I_{\phantom{s}\alpha}=(P^I_{\phantom{a}0},P^I_{\phantom{a}r})=(\Omega^I_{\phantom{a}0},\Omega^I_{\phantom{a}r}).
\end{equation}
\indent The structure constants of the full $SU(2)_R\times SU(2)$
gauge group $f^\Lambda_{\phantom{as}\Pi\Sigma}$ will be split into
$\epsilon_{rst}$ and $C_{IJK}=\epsilon_{IJK}$ for the two factors
$SU(2)_R$ and $SU(2)$, respectively. There are accordingly two
coupling constants denoted by $g_1$ and $g_2$.
\\
\indent In order to parametrize scalar fields described by the
$SO(4,3)/SO(4)\times SO(3)$ coset, we introduce basis elements of
$7\times 7$ matrices by
\begin{equation}
(e^{\Lambda \Sigma})_{\Gamma \Pi}=\delta_{\Lambda
\Gamma}\delta_{\Sigma \Pi},\qquad \Lambda, \Sigma,\Gamma,
\Pi=0,\ldots ,6\, .
\end{equation}
The $SO(4)$, $SU(2)_R$, $SU(2)$ and non-compact generators
$Y_{\alpha I}$ of $SO(4,3)$ are then given by
\begin{eqnarray}
SO(4):\qquad
J^{\alpha\beta}&=&e^{\beta,\alpha}-e^{\alpha,\beta},\qquad \alpha,\beta=0,1,2,3,\nonumber \\
SU(2)_{R}:\qquad J_1^{rs}&=&e^{s,r}-e^{r,s},\qquad r,s=1,2,3,\nonumber \\
SU(2):\qquad J_2^{IJ}&=&e^{J+3,I+3}-e^{I+3,J+3},\qquad I,J=1,2,3,\nonumber \\
Y_{\alpha I}&=&e^{\alpha,I+3}+e^{I+3,\alpha}\, .
\end{eqnarray}
\indent In this paper, we are not interested in solutions with
non-zero two-form field. We therefore set $B_{\mu\nu}=0$ from now
on. The bosonic Lagrangian involving only the metric, vectors and
scalar fields is given by \cite{F4SUGRA2}
\begin{equation}
\mathcal{L}=\frac{1}{4}eR-e\pd_\mu \sigma\pd^\mu \sigma
-\frac{1}{4}eP_{I\alpha\mu}P^{I\alpha\mu}-\frac{1}{8}ee^{-2\sigma}\mc{N}_{\Lambda\Sigma}F^\Lambda_{\mu\nu}
F^{\Sigma\mu\nu}-eV
\end{equation}
where $e=\sqrt{-g}$. We have written the scalar kinetic term in term
of $P^{I\alpha}_\mu=P^{I\alpha}_i\pd_\mu\phi^i$, $i=1,\ldots, 12$.
The explicit form of the scalar potential is given by
\begin{eqnarray}
V&=&-e^{2\sigma}\left[\frac{1}{36}A^2+\frac{1}{4}B^iB_i+\frac{1}{4}\left(C^I_{\phantom{s}t}C_{It}+4D^I_{\phantom{s}t}D_{It}\right)\right]
+m^2e^{-6\sigma}\mc{N}_{00}\nonumber \\
& &-me^{-2\sigma}\left[\frac{2}{3}AL_{00}-2B^iL_{0i}\right]
\end{eqnarray}
where $\mc{N}_{00}$ is the $00$ component of the scalar matrix
$\mc{N}_{\Lambda\Sigma}$ defined by
\begin{equation}
\mc{N}_{\Lambda\Sigma}=L^{\phantom{as}0}_\Lambda
(L^{-1})_{0\Sigma}+L^{\phantom{as}i}_\Lambda
(L^{-1})_{i\Sigma}-L^{\phantom{as}I}_\Lambda (L^{-1})_{I\Sigma}\, .
\end{equation}
Various quantities appearing in the scalar potential and the
supersymmetry transformations given below are defined as follow
\begin{eqnarray}
A&=&\epsilon^{rst}K_{rst},\qquad B^i=\epsilon^{ijk}K_{jk0},\\
C^{\phantom{ts}t}_I&=&\epsilon^{trs}K_{rIs},\qquad D_{It}=K_{0It}
\end{eqnarray}
where
\begin{eqnarray}
K_{rst}&=&g_1\epsilon_{lmn}L^l_{\phantom{r}r}(L^{-1})_s^{\phantom{s}m}L_{\phantom{s}t}^n+
g_2C_{IJK}L^I_{\phantom{r}r}(L^{-1})_s^{\phantom{s}J}L_{\phantom{s}t}^K,\nonumber
\\
K_{rs0}&=&g_1\epsilon_{lmn}L^l_{\phantom{r}r}(L^{-1})_s^{\phantom{s}m}L_{\phantom{s}0}^n+
g_2C_{IJK}L^I_{\phantom{r}r}(L^{-1})_s^{\phantom{s}J}L_{\phantom{s}0}^K,\nonumber
\\
K_{rIt}&=&g_1\epsilon_{lmn}L^l_{\phantom{r}r}(L^{-1})_I^{\phantom{s}m}L_{\phantom{s}t}^n+
g_2C_{IJK}L^I_{\phantom{r}r}(L^{-1})_I^{\phantom{s}J}L_{\phantom{s}t}^K,\nonumber
\\
K_{0It}&=&g_1\epsilon_{lmn}L^l_{\phantom{r}0}(L^{-1})_I^{\phantom{s}m}L_{\phantom{s}t}^n+
g_2C_{IJK}L^I_{\phantom{r}0}(L^{-1})_I^{\phantom{s}J}L_{\phantom{s}t}^K\,
.
\end{eqnarray}
\indent Finally, we need supersymmetry transformations of $\chi^A$,
$\lambda^I_A$ and $\psi^A_\mu$ to find supersymmetric bosonic
solutions. These transformation rules with vanishing $B_{\mu\nu}$
field are given by
\begin{eqnarray}
\delta\psi_{\mu
A}&=&D_\mu\epsilon_A-\frac{1}{24}\left(Ae^\sigma+6me^{-3\sigma}(L^{-1})_{00}\right)\epsilon_{AB}\gamma_\mu\epsilon^B\nonumber
\\
& &-\frac{1}{8}
\left(B_te^\sigma-2me^{-3\sigma}(L^{-1})_{t0}\right)\gamma^7\sigma^t_{AB}\gamma_\mu\epsilon^B\nonumber \\
&
&+\frac{i}{16}e^{-\sigma}\left[\epsilon_{AB}(L^{-1})_{0\Lambda}\gamma_7+\sigma^r_{AB}(L^{-1})_{r\Lambda}\right]
F^\Lambda_{\nu\lambda}(\gamma_\mu^{\phantom{s}\nu\lambda}
-6\delta^\nu_\mu\gamma^\lambda)\epsilon^B,\label{delta_psi}\\
\delta\chi_A&=&\frac{1}{2}\gamma^\mu\pd_\mu\sigma\epsilon_{AB}\epsilon^B+\frac{1}{24}
\left[Ae^\sigma-18me^{-3\sigma}(L^{-1})_{00}\right]\epsilon_{AB}\epsilon^B\nonumber
\\
& &-\frac{1}{8}
\left[B_te^\sigma+6me^{-3\sigma}(L^{-1})_{t0}\right]\gamma^7\sigma^t_{AB}\epsilon^B\nonumber
\\
&
&+\frac{i}{16}e^{-\sigma}\left[\epsilon_{AB}(L^{-1})_{0\Lambda}\gamma_7-
\sigma^r_{AB}(L^{-1})_{r\Lambda}\right]F^\Lambda_{\mu\nu}\gamma^{\mu\nu}\epsilon^B,
\label{delta_chi}\\
\delta
\lambda^{I}_A&=&P^I_{ri}\gamma^\mu\pd_\mu\phi^i\sigma^{r}_{\phantom{s}AB}\epsilon^B+P^I_{0i}
\gamma^7\gamma^\mu\pd_\mu\phi^i\epsilon_{AB}\epsilon^B-\left(2i\gamma^7D^I_{\phantom{s}t}+C^I_{\phantom{s}t}\right)
e^\sigma\sigma^t_{AB}\epsilon^B \nonumber
\\
& &+2me^{-3\sigma}(L^{-1})^I_{\phantom{ss}0}
\gamma^7\epsilon_{AB}\epsilon^B-\frac{i}{2}e^{-\sigma}(L^{-1})^I_{\phantom{s}\Lambda}F^\Lambda_{\mu\nu}
\gamma^{\mu\nu}\epsilon_{A}\label{delta_lambda}
\end{eqnarray}
where $\sigma^{tC}_{\phantom{sd}B}$ are usual Pauli matrices, and
$\epsilon_{AB}=-\epsilon_{BA}$. In our convention, the space-time
gamma matrices $\gamma^a$ satisfy
\begin{equation}
\{\gamma^a,\gamma^b\}=2\eta^{ab},\qquad
\eta^{ab}=\textrm{diag}(-1,1,1,1,1,1),
\end{equation}
and $\gamma^7=\gamma^0\gamma^1\gamma^2\gamma^3\gamma^4\gamma^5$ with
$\gamma_7^2=\mathbf{1}$. The covariant derivative of $\epsilon_A$ is
given by
\begin{equation}
D_\mu \epsilon_A=\pd_\mu
\epsilon_A+\frac{1}{4}\omega_\mu^{ab}\gamma_{ab}\epsilon_A+\frac{i}{2}\sigma_{rAB}
\left[\frac{1}{2}\epsilon^{rst}\Omega_{\mu st}-i\gamma_7
\Omega_{\mu r0}\right]\epsilon^B\, .
\end{equation}
It should be noted that due to some difference in conventions, the
above supersymmetry transformations do not coincide with those of
the pure $F(4)$ gauged supergravity given in \cite{F4_Romans} when
all of the fields in the vector multiplets are set to zero. However,
it can be verified that the transformation rules in \cite{F4_Romans}
are recovered by using the identifications
\begin{equation}
\gamma^\mu\rightarrow \gamma_7\gamma^\mu\qquad \textrm{and}\qquad
\chi_A\rightarrow \gamma_7 \chi_A\, .
\end{equation}
\indent The $SU(2)_R\times SU(2)$ gauged supergravity admits
maximally supersymmetric $AdS_6$ critical points when $m\neq 0$. One
of them is the trivial critical point at which all scalars vanish
after setting $g_1=3m$. This critical point preserves the full
$SU(2)_R\times SU(2)$ symmetry and should be dual to the
five-dimensional SCFT with global symmetry $E_1\sim SU(2)$.
Furthermore, at the vacuum, the $U(1)$ gauge field $A^0$ will be
eaten by the two-form field resulting in a massive $B_{\mu\nu}$
field. Another supersymmetric $AdS_6$ critical point preserves only
the diagonal subgroup $SU(2)_{\textrm{diag}}\subset SU(2)_R\times
SU(2)$. This critical point has been mistakenly identified as a
stable non-supersymmetric $AdS_6$ in \cite{F4_flow}, see also the
associated erratum.
\\
\indent Actually, the non-trivial supersymmetric critical point can
also be seen from the BPS equations studied in \cite{5DSYM_from_F4},
but that paper mainly considers RG flows from five-dimensional SCFTs
corresponding to the trivial $AdS_6$ critical point to non-conformal
field theories in the IR. We give the analysis of these two
supersymmetric $AdS_6$ critical points in the appendix together with
an analytic RG flow between them. This flow solution have already
been studied numerically in \cite{F4_flow}. The critical points and
the flow solution are similar to the corresponding solutions in the
half-maximal gauged supergravity with $SO(4)$ gauge group in seven
dimensions studied in \cite{7D_flow}.
\section{$AdS_4$ critical points}\label{AdS4}
In this section, we consider solutions of the form $AdS_4\times S^2$
or $AdS_4\times H^2$ with $S^2$ and $H^2$ being a two-sphere and a
two-dimensional hyperbolic space, respectively. The metric takes the
form of
\begin{equation}
ds^2=e^{2F}dx^2_{1,2}+e^{2G}(d\theta^2+\sin^2\theta d\phi^2)+dr^2
\end{equation}
for the $S^2$ case and
\begin{equation}
ds^2=e^{2F}dx^2_{1,2}+\frac{e^{2G}}{y^2}(dx^2+dy^2)+dr^2
\end{equation}
for the $H^2$ case. In both cases, the warp factors $F$ and $G$ are
functions only of $r$.
\\
\indent The non-vanishing spin connections of the above metrics are
given respectively by
\begin{eqnarray}
\omega^{\hat{\phi}}_{\phantom{s}\hat{\theta}} &=&e^{-G}\cot\theta
e^{\hat{\phi}},\qquad
\omega^{\hat{\phi}}_{\phantom{s}\hat{r}}=G'e^{\hat{\phi}},\nonumber \\
\omega^{\hat{\theta}}_{\phantom{s}\hat{r}}&=&G'e^{\hat{\theta}},\qquad
\omega^{\hat{\mu}}_{\phantom{s}\hat{r}}=F'e^{\hat{\mu}}
\end{eqnarray}
and
\begin{eqnarray}
\omega^{\hat{x}}_{\phantom{s}\hat{r}}&=&G'e^{\hat{x}},\qquad
\omega^{\hat{y}}_{\phantom{s}\hat{r}}=G'e^{\hat{y}},\nonumber \\
\omega^{\hat{\mu}}_{\phantom{s}\hat{r}}&=&F'e^{\hat{\mu}},\qquad
\omega^{\hat{x}}_{\phantom{s}\hat{y}}=-e^{-G(r)}e^{\hat{x}}\label{spin_connection_H2}
\end{eqnarray}
where $'$ denotes the $r$-derivative.

\subsection{$N=2$ three-dimensional SCFTs with $SO(2)\times SO(2)$ symmetry}
To find supersymmetric solutions of the form $AdS_4\times
\Sigma_{2}$ with $SO(2)\times SO(2)$ symmetry, we turn on
$SO(2)\times SO(2)$ gauge fields such that the spin connection along
$\Sigma$ is canceled. In the present case, there are six gauge
fields $(A^r,A^I)$ corresponding to $SU(2)_R\times SU(2)$ gauge
group. We will turn on the following $SO(2)\times SO(2)$ gauge
fields
\begin{equation}
A^3=a\cos \theta d\phi\qquad \textrm{and}\qquad A^6=b\cos \theta
d\phi
\end{equation}
for the $S^2$ case and
\begin{equation}
A^3=\frac{a}{y}dx\qquad\textrm{and} \qquad A^6=\frac{b}{y}dx
\end{equation}
for the $H^2$ case. To avoid confusion, we have given the gauge
fields using the index $\Lambda=0,1,\ldots, 6$.
\\
\indent $A^3$ will appear in the covariant derivative of
$\epsilon^A$ since it is part of the $SU(2)_R$ gauge fields. We
choose this particular form of the gauge field to cancel the spin
connection on $\Sigma_2$. Accordingly, the Killing spinors
corresponding to unbroken supersymmetry will be constant spinors on
$\Sigma_2$ provided that we impose the twist condition
\begin{equation} ag_1=1
\end{equation}
and a set of projection conditions given below.
\\
\indent There are two scalars which are singlet under $SO(2)\times
SO(2)$ generated by $J_1^{12}$ and $J_2^{12}$. The
$SO(4,3)/SO(4)\times SO(3)$ coset representative can be written in
terms of these scalars as
\begin{equation}
L=e^{\phi_1 Y_{03}}e^{\phi_2 Y_{33}}\, .
\end{equation}
\indent Imposing the projection conditions
\begin{equation}
\gamma_{\hat{r}}\epsilon_A=\epsilon_A,\qquad
\gamma_7\epsilon^A=\delta^A_{B}\epsilon^B,\qquad
\gamma^{\hat{\phi}\hat{\theta}}\epsilon_A=i\sigma_{3AB}\epsilon^B,
\end{equation}
for the $S^2$ case or
\begin{equation}
\gamma_{\hat{r}}\epsilon_A=\epsilon_A,\qquad
\gamma_7\epsilon^A=\delta^A_{B}\epsilon^B,\qquad
\gamma^{\hat{x}\hat{y}}\epsilon_A=-i\sigma_{3AB}\epsilon^B,
\end{equation}
for the $H^2$ case, we find that consistency of the BPS equations from $\delta \psi_{A \mu}$, for $\mu=0,1,2$, requires $\phi_1=0$. Setting $\phi_1=0$, we obtain the following BPS equations
\begin{eqnarray}
\phi_2'&=&-\frac{1}{4}e^{-\sigma-\phi_2-2G}\left[2\lambda
b(1+e^{2\phi_2})
+2(1-e^{2\phi_2})(\lambda
a+2g_1e^{2\sigma+2G})\right],\\
\sigma'&=&\frac{1}{8}e^{-3\sigma-\phi_2-2G}\left[\lambda
ae^{2\sigma}(1+e^{2\phi_2})-\lambda be^{2\sigma}(2e^{\phi_2}-1) \right .\nonumber \\ 
& &\left. -2e^{2G}[g_1e^{4\sigma}(1+e^{2\phi_2})-6me^{\phi_2}]\right],\\
G'&=&\frac{1}{8}e^{-3\sigma-\phi_2-2G}\left[3\lambda
ae^{2\sigma}(1+e^{2\phi_2})-3\lambda
be^{2\sigma}(e^{2\phi_2}-1)\right. \nonumber
\\
&
&\left.+2e^{2G}[g_1e^{4\sigma}(1+e^{2\phi_2})+2me^{\phi_2}]\right],\\
F'&=&\frac{1}{8}e^{-3\sigma-\phi_2-2G}\left[\lambda
be^{2\sigma}(e^{2\phi_2}-1)-\lambda
ae^{2\sigma}(1+e^{2\phi_2})\right. \nonumber
\\
&
&\left.+2e^{2G}[g_1e^{4\sigma}(1+e^{2\phi_2})+2me^{\phi_2}]\right]
\end{eqnarray}
where $\lambda=1$ and $\lambda=-1$ for $S^2$ and $H^2$ cases,
respectively.
\\
\indent We look for fixed point solutions satisfying
$G'=\sigma'=\phi_2'=0$ and $F\sim r$. The $\gamma_{\hat{r}}$ projector is not necessary for constant scalars since $\gamma_{\hat{r}}$ only appears with the
$r$-derivative. The BPS equations are automatically satisfied by the
fixed point solutions without imposing the $\gamma_{\hat{r}}$
projector. Furthermore, with $\phi_1=0$, the $\gamma_7$ projection
is not needed. Therefore, the $AdS_4$ fixed points will preserve
half of the original supersymmetry corresponding to eight
supercharges or $N=2$ superconformal symmetry in three dimensions.
\\
\indent The explicit form of $AdS_4$ critical point
is given by
\begin{eqnarray}
\phi_2&=&\frac{1}{2}\ln\left[\frac{3b\pm
\sqrt{a^2+8b^2}}{b-a}\right],\nonumber \\
\sigma&=&\frac{1}{8}\ln\left[\frac{m^2(b-a)(a+4b\mp
\sqrt{a^2+8b^2})^2}
{4b^2g_1^2(3b\mp \sqrt{a^2+8b^2})}\right], \nonumber \\
G&=&\frac{1}{8}\ln\left[\frac{b^2(b-a)^3(3b\mp
\sqrt{a^2+8b^2})(a+4b\mp\sqrt{a^2+8b^2})^2}
{4g_1^2m^2(a+2b\mp\sqrt{a^2+8b^2})^4}\right],\nonumber \\
L_{AdS_4}&=&\frac{1}{2m}\left[\frac{(b-a)m^2(a+4b\pm\sqrt{a^2+8b^2})^2}{4b^2g_1^2(3b\pm\sqrt{a^2+8b^2})}
\right]^{\frac{3}{8}}.
\end{eqnarray}
In the above equations, we have given a solution in the $S^2$ case
for definiteness. A similar solution in the $H^2$ case can be
obtained by replacing $(a,b)$ by $(-a,-b)$ in the above solution.
For $a<0$, the solution is valid provided that $b<a$ or $b>-a$. When
$a>0$, we have a real solution for $b<a$ or $b>a$. It can be checked
that there exist both $AdS_4\times S^2$ and $AdS_4\times H^2$ fixed
points.
\\
\indent As an example, we give some $AdS_4$ solutions with a
particular value of $b=2a$ as follow:
\begin{eqnarray}
AdS_4\times S^2:\quad \phi_2&=&\frac{1}{2}\ln(6+\sqrt{33}),\qquad
\sigma=\frac{1}{4}\ln\left[\frac{(9+\sqrt{33})m}{4\sqrt{6+\sqrt{33}}g+1}\right],\nonumber
\\
G&=&\frac{1}{8}\ln\left[\frac{6a^4(213+37\sqrt{33})}{g_1^2m^2(5+\sqrt{33})^4}\right]
\end{eqnarray}
and
\begin{eqnarray}
AdS_4\times H^2:\quad
\phi_2&=&\frac{1}{2}\ln(2+\sqrt{\frac{1}{3}}),\qquad
\sigma=\frac{1}{4}\ln\left[\frac{(7\sqrt{3}+3\sqrt{11})m}{4\sqrt{6+\sqrt{33}}g+1}\right],\nonumber
\\
G&=&\frac{1}{8}\ln\left[\frac{59a^4(477+83\sqrt{33})}{g_1^2m^2(3+\sqrt{33})^4}\right].
\end{eqnarray}
It can also be readily verified that, by making a truncation
$\phi_2=0$ and $b=0$, we find only $AdS_4\times H^2$ solution in
agreement with the results of \cite{F4_nunez}. It should also be
pointed out that the solutions are similar to the ones obtained in
seven-dimensional gauged supergravity studied in \cite{AdS5_4_N2_7D}
and \cite{Cucu1,Cucu2}. It is also possible to find a numerical RG
flow solution interpolating between $SU(2)\times SU(2)$ $AdS_6$
critical point \eqref{SO4_AdS6} to one of these $AdS_4$ critical
points, but we will not give it here.

\subsection{$N=2$ three-dimensional SCFTs with $SO(2)$ symmetry}
We now consider $AdS_4$ solutions that can be connected to the
$AdS_6$ critical point with $SU(2)_{\textrm{diag}}$ symmetry
\eqref{SO3_AdS6} by an interpolating domain wall solution. In this
case, there can be RG flows from $AdS_6$ critical point in
\eqref{SO3_AdS6} to three-dimensional SCFTs in the IR or even a flow
from $AdS_6$ critical point \eqref{SO4_AdS6} to critical point
\eqref{SO3_AdS6} and then to the $AdS_4$ points.
\\
\indent We look for solutions preserving $SO(2)_{\textrm{diag}}$
subgroup of $SO(2)\times SO(2)$ generated by $J_1^{12}+J_2^{12}$.
The gauge fields for the $S^2$ and $H^2$ cases are then given
respectively by
\begin{equation}
A^3=a\cos \theta d\phi\qquad \textrm{and} \qquad
A^6=\frac{g_1}{g_2}A^3
\end{equation}
and
\begin{equation}
A^3=\frac{a}{y} dx\qquad \textrm{and} \qquad
A^6=\frac{g_1}{g_2}A^3\, .
\end{equation}
\indent There are four $SO(2)_{\textrm{diag}}$ singlet scalars with
the scalar coset representative given by
\begin{equation}
L=e^{\phi_1(Y_{11}+Y_{22})}e^{\phi_2 Y_{33}}e^{\phi_3
Y_{03}}e^{\phi_4 (Y_{12}-Y_{21})}\, .
\end{equation}
By using similar projection conditions and the relation $g_1a=1$ as
in the previous case, we find that consistency of the BPS equations requires $\phi_3=0$. Moreover, as in the previous case, the $\gamma_7$ projector is
irrelevant when $\phi_3=0$. Therefore, the fixed point solutions
will also preserve eight supercharges. The corresponding BPS equations are given by
\begin{eqnarray}
\phi_4'&=&\frac{1}{8}e^{\sigma-2\phi_1-\phi_2-2\phi_4}(1+e^{4\phi_1})(1-e^{4\phi_4})
(g_1+g_1e^{2\phi_2}+g_2-g_2e^{2\phi_2}),\\
\phi_1'&=&\frac{e^{\sigma-2\phi_1-\phi_2+2\phi_4}(1-e^{4\phi_1})}{2(1+e^{4\phi_4})}
(g_1+g_1e^{2\phi_2}+g_2-g_2e^{2\phi_2}),\\
\phi_2'&=&-\frac{1}{8g_2}e^{-\sigma-2\phi_1-\phi_2-2\phi_4-2G}\left[-4\lambda
ae^{2\phi_1+2\phi_4}(g_1+g_1e^{2\phi_2}+g_2-g_2e^{2\phi_2})\right.\nonumber
\\
&
&-g_2e^{2\sigma+2G}\left[g_1(1-e^{2\phi_2})(1+e^{4\phi_1}+e^{4\phi_4}+4e^{2\phi_1+2\phi_4}
+e^{4\phi_1+4\phi_4})\right.\nonumber \\
&
&\left.\left.+g_2(1+e^{2\phi_2})(1+e^{4\phi_1}+e^{4\phi_4}-4e^{2\phi_1+2\phi_4}+e^{4\phi_1+4\phi_4})
\right]\right],\\
\sigma'&=&\frac{1}{32}e^{-3\sigma-2\phi_1-\phi_2-2\phi_4-2G}\left[48me^{2\phi_1+\phi_2+2\phi_4+2G}\right.\nonumber
\\
&
&-(g_1-g_2)e^{4\sigma+2\phi_2+2G}(1+e^{4\phi_1}+4e^{2\phi_1+2\phi_4-2\phi_2}+e^{4\phi_4}
+e^{4\phi_1+4\phi_4})\nonumber \\
&
&-(g_1+g_2)e^{4\sigma+2G}(1+e^{4\phi_4}+e^{4\phi_4}+e^{4\phi_1+4\phi_4}+4e^{2\phi_1+2\phi_2+2\phi_4})\nonumber
\\
& &\left. +\frac{4\lambda
a}{g_2}e^{2\sigma+2\phi_1+2\phi_4}(g_1-g_1e^{2\phi_2}+g_2+g_2e^{2\phi_2})\right],\\
G'&=&\frac{1}{32}e^{-3\sigma-2\phi_1-\phi_2-2\phi_4-2G}\left[16me^{2\phi_1+\phi_2+2\phi_4+2G}\right.\nonumber
\\
&
&+(g_1-g_2)e^{4\sigma+2\phi_2+2G}(1+e^{4\phi_1}+4e^{2\phi_1+2\phi_4-2\phi_2}+e^{4\phi_4}
+e^{4\phi_1+4\phi_4})\nonumber \\
&
&+(g_1+g_2)e^{4\sigma+2G}(1+e^{4\phi_4}+e^{4\phi_4}+e^{4\phi_1+4\phi_4}+4e^{2\phi_1+2\phi_2+2\phi_4})\nonumber
\\
& &\left. +\frac{12\lambda
a}{g_2}e^{2\sigma+2\phi_1+2\phi_4}(g_1-g_1e^{2\phi_2}+g_2+g_2e^{2\phi_2})\right],\\
F'&=&\frac{1}{32}e^{-3\sigma-2\phi_1-\phi_2-2\phi_4-2G}\left[16me^{2\phi_1+\phi_2+2\phi_4+2G}\right.\nonumber
\\
&
&+(g_1-g_2)e^{4\sigma+2\phi_2+2G}(1+e^{4\phi_1}+4e^{2\phi_1+2\phi_4-2\phi_2}+e^{4\phi_4}
+e^{4\phi_1+4\phi_4})\nonumber \\
&
&+(g_1+g_2)e^{4\sigma+2G}(1+e^{4\phi_4}+e^{4\phi_4}+e^{4\phi_1+4\phi_4}+4e^{2\phi_1+2\phi_2+2\phi_4})\nonumber
\\
& &\left. -\frac{4\lambda
a}{g_2}e^{2\sigma+2\phi_1+2\phi_4}(g_1-g_1e^{2\phi_2}+g_2+g_2e^{2\phi_2})\right]
\end{eqnarray}
where, as in the previous case, $\lambda=\pm 1$ for $S^2$ and $H^2$,
respectively. When $a=0$, $\phi_2=\phi_1=\phi$ and $\phi_4=0$, we
recover the BPS equations \eqref{6D_eq1}, \eqref{6D_eq2} and
\eqref{6D_eq3} given in the appendix.
\\
\indent We begin with a simple solution for $\phi_4=0$. There are
two possibilities for the critical points to occur depending on the
values of the coupling constants $g_1$ and $g_2$. The critical point
that can be connected to the $AdS_6$ critical point \eqref{SO3_AdS6}
for which $g_2<-g_1$ and $g_1=3m>0$ is given by
\begin{eqnarray}
\phi_2&=&\frac{1}{2}\ln\left[\frac{g_2+g_1}{g_2-g_1}\right],\qquad
\phi_1=\pm
\frac{1}{2}\ln\left[\frac{g_2+g_1}{g_2-g_1}\right],\nonumber \\
\sigma&=&\frac{1}{4}\ln\left[-\frac{2m\sqrt{g_2^2-g_1^2}}{g_1g_2}\right],\qquad
G=\frac{1}{2}\ln\left[-\frac{a}{g_2m}\sqrt{-\frac{m(g_2^2-g_1^2)^{\frac{3}{2}}}{2g_1g_2}}\right],\nonumber
\\
L_{AdS_4}&=&\frac{1}{2m}\left[-\frac{2m\sqrt{g_2^2-g_1^2}}{g_1g_2}\right]^{\frac{3}{4}}\,
.
\end{eqnarray}
It can be verified that, in this case, only $AdS_4\times H^2$
solutions are possible. The other possibility with positive $g_2$
however does not give any real solutions.
\\
\indent For a particular value of $g_2=g_1$, there is an
$AdS_4\times H^2$ solution given by
\begin{eqnarray}
\phi_2&=&-\frac{1}{2}\ln 3,\qquad
\phi_1=0,\nonumber \\
\sigma&=&\frac{1}{8}\ln 3+\frac{1}{4}\ln\left[\frac{m}{g_1}\right],\qquad
G=-\frac{1}{2}\ln\left[\frac{2g_1}{3^{\frac{3}{4}}a}\sqrt{\frac{m}{g_1}}\right].
\end{eqnarray}
\indent For non-zero $\phi_4$, there is a class of solutions parametrized by $\phi_4$. The explicit form of these solutions for $g_2<-g_1$ and $g_1=3m>0$ is given by
\begin{eqnarray}
\phi_2&=&\frac{1}{2}\ln\left[\frac{g_2+g_1}{g_2-g_1}\right],\qquad \sigma=\frac{1}{4}\ln\left[-\frac{2m\sqrt{g_2^2-g_1^2}}{g_1g_2}\right],\nonumber \\
G&=&\frac{1}{2}\ln\left[-\frac{a}{g_2m}\sqrt{-\frac{m(g_2^2-g_1^2)^{\frac{3}{2}}}{2g_1g_2}}\right],\nonumber \\
\phi_1&=&\frac{1}{2}\ln\left[\frac{2e^{2\phi_4}(g_1^2+g_2^2)+\sqrt{4e^{4\phi_4}(g_1^2+g_2^2)^2-(1+e^{4\phi_4})^2
(g_2^2-g_1^2)^2}}{(g_2^2-g_1^2)(1+e^{4\phi_4})}\right].\quad
\end{eqnarray}
This solution is also $AdS_4\times H^2$ as in the previous case and
valid for
\begin{equation}
\frac{(g_2+g_1)^2}{(g_2-g_1)^2}\leq e^{4\phi_4} \leq \frac{(g_2-g_1)^2}{(g_2+g_1)^2}\, .
\end{equation}
\indent In this case, the scalar $\phi_4$ is not determined by the
BPS equations. Consider this critical point to be an IR fixed point
of the five-dimensional SCFTs corresponding to the $AdS_6$ critical
point \eqref{SO3_AdS6}, we can see that $\phi_4$ corresponds to a
marginal deformation since $\phi_4$ is massless as can be seen from
the scalar masses given in \cite{F4_flow}. It should also be noted
that when the symmetry is reduced to $SO(2)$, only solutions with a
hyperbolic space are possible. This is similar to the
seven-dimensional results studied in
\cite{AdS5_4_N2_7D,Cucu1,Cucu2}.

\section{$AdS_3$ critical points}\label{AdS3}
We now look for a gravity dual of five-dimensional SCFTs
compactified on a three-manifold $\Sigma_3$ which can be $S^3$ or
$H^3$. The IR effective theories would be two-dimensional field
theories. We particularly look for the gravity solutions
corresponding to conformal field theories in the IR, so the gravity
solutions will take the form of $AdS_3\times \Sigma_3$.
\\
\indent The metrics and the associated spin connections for each case are given by
\begin{equation}
ds^2_7=e^{2F}dx^2_{1,2}+dr^2+e^{2G}\left[d\psi^2+\sin^2\psi
(d\theta^2+\sin^2\theta d\phi^2)\right]
\end{equation}
for $\Sigma_3=S^3$ with the spin connections
\begin{eqnarray}
\omega^{\hat{\mu}}_{\phantom{s}\hat{r}}&=&F'e^{\hat{\mu}},\qquad \omega^{\hat{\psi}}_{\phantom{s}\hat{r}}=G'e^{\hat{\psi}},\qquad
\omega^{\hat{\theta}}_{\phantom{s}\hat{r}}=G'e^{\hat{\theta}},\nonumber \\
\omega^{\hat{\phi}}_{\phantom{s}\hat{r}}&=&G'e^{\hat{\phi}},\qquad \omega^{\hat{\phi}}_{\phantom{s}\hat{\theta}}=e^{-G}\frac{\cot\theta}{\sin \psi}e^{\hat{\phi}},\nonumber \\
\omega^{\hat{\phi}}_{\phantom{s}\hat{\psi}}&=&e^{-G}\cot\psi e^{\hat{\phi}},\qquad
\omega^{\hat{\theta}}_{\phantom{s}\hat{\psi}}=e^{-G}\cot\psi e^{\hat{\theta}}
\end{eqnarray}
and
\begin{equation}
ds^2_7=e^{2F}dx^2_{1,2}+dr^2+\frac{e^{2G}}{y^2}(dx^2+dy^2+dz^2)\label{AdS4H3_metric}
\end{equation}
for $\Sigma_3=H^3$ with the spin connections given by
\begin{eqnarray}
\omega^{\hat{z}}_{\phantom{s}\hat{r}}&=&G'e^{\hat{z}},\qquad \omega^{\hat{y}}_{\phantom{s}\hat{r}}=G'e^{\hat{y}},\qquad \omega^{\hat{x}}_{\phantom{s}\hat{r}}=G'e^{\hat{x}},\nonumber \\
\omega^{\hat{x}}_{\phantom{s}\hat{y}}&=&-e^{-G}e^{\hat{x}},\qquad \omega^{\hat{z}}_{\phantom{s}\hat{y}}=-e^{-G}e^{\hat{z}},\qquad \omega^{\hat{\mu}}_{\phantom{s}\hat{r}}=F'e^{\hat{\mu}}\, .
\end{eqnarray}

\subsection{$N=(1,1)$ two-dimensional SCFTs with $SO(3)$ symmetry}
We first look at solutions preserving $SO(3)_{\textrm{diag}}\subset
SO(3)_R\times SO(3)$ symmetry. The $SO(4,3)/SO(4)\times SO(3)$ coset
representative is given in \eqref{L_SO3D}. We then proceed by
turning on $SO(3)_{\textrm{diag}}$ gauge fields to cancel the spin
connections on $\Sigma_3$ as in the previous section.
\\
\indent For the $S^3$ case, we choose the $SU(2)_R$ gauge fields to be
\begin{equation}
A^1=a\cos\psi \sin\theta d\phi,\qquad A^2=b\cos\theta d\phi ,\qquad
A^3=c\cos \psi d\theta
\end{equation}
while, for the $H^3$ case, they are given by
\begin{equation}
A^1=\frac{a}{y}dz,\qquad A^2=0,\qquad A^3=\frac{b}{y}dx\, .
\end{equation}
In both cases, the $SO(3)$ gauge fields are related to the $SU(2)_R$
gauge fields by
\begin{equation}
A^I=\frac{g_1}{g_2}A^r\, .
\end{equation}
The two sets of gauge fields implement the $SO(3)_{\textrm{diag}}$
gauge fields. Furthermore, the twist condition implies $a=b=c$ and
$g_1a=1$.
\\
\indent Using the following projectors
\begin{eqnarray}
H^3:\qquad \gamma_{\hat{r}}\epsilon_A&=&\epsilon_A,\qquad \gamma_{\hat{x}\hat{y}}\epsilon_A=-i\sigma^3_{AB}\epsilon^B,\nonumber \\
\gamma_{\hat{z}\hat{y}}\epsilon_A&=&-i\sigma^1_{AB}\epsilon^B,\qquad
\gamma_{\hat{x}\hat{z}}\epsilon_A=-i\sigma^2_{AB}\epsilon^B,\label{S3_projetor}\\
S^3:\qquad \gamma_{\hat{r}}\epsilon_A&=&\epsilon_A,\qquad \gamma_{\hat{\theta}\hat{\psi}}\epsilon_A=i\sigma^3_{AB}\epsilon^B,\nonumber \\
\gamma_{\hat{\phi}\hat{\psi}}\epsilon_A&=&i\sigma^1_{AB}\epsilon^B,\qquad
\gamma_{\hat{\phi}\hat{\theta}}\epsilon_A=i\sigma^2_{AB}\epsilon^B,\label{H3_projetor}
\end{eqnarray}
we obtain the BPS equations
\begin{eqnarray}
\phi'&=&\frac{1}{4g_2}e^{-\sigma-3\phi-2G}[(1+e^{2\phi})g_1+(1-e^{2\phi})g_2][2\lambda ae^{2\phi}
+g_2e^{2\phi+2G}(1-e^{4\phi})],\qquad\\
\sigma'&=&\frac{3}{8g_2}\lambda ae^{-\sigma-\phi-2G}[(1-e^{2\phi})g_1+(1+e^{2\phi})g_2]-\frac{3}{2}me^{-3\sigma}\nonumber \\
& &-\frac{1}{16}[(1+e^{2\phi})^3g_1+(1-e^{2\phi})^3g_2],\\
G'&=&\frac{1}{16}e^{\sigma-3\phi}[(1+e^{2\phi})^3g_1+(1-e^{2\phi})^3g_2]+\frac{1}{2}me^{-3\sigma}\nonumber \\
& &+\frac{5}{8g_2}\lambda ae^{-\sigma-\phi-2G}[(1-e^{2\phi})g_1+(1+e^{2\phi})g_2],\\
F'&=&\frac{1}{16}e^{\sigma-3\phi}[(1+e^{2\phi})^3g_1+(1-e^{2\phi})^3g_2]+\frac{1}{2}me^{-3\sigma}\nonumber \\
& &-\frac{3}{8g_2}\lambda ae^{-\sigma-\phi-2G}[(1-e^{2\phi})g_1+(1+e^{2\phi})g_2]
\end{eqnarray}
where $\lambda=\pm 1$ for $S^3$ and $H^3$ as in the previous cases.
Note also that, in both cases, the last projector is not independent
from the second and the third ones. Therefore, fixed point solutions
will preserve four supercharges or equivalently $N=(1,1)$
superconformal symmetry in two dimensions. The analysis of unbroken
supersymmetry can be done in a similar manner to that given in
\cite{F4_nunez}.
\\
\indent We begin with a simple fixed point solution with $g_2=g_1$.
In this case, only $AdS_3\times H^3$ solution exists and is given by
\begin{eqnarray}
\phi&=&-\frac{1}{4}\ln 3,\qquad \sigma=\frac{1}{16}\ln 3-\frac{1}{4}\ln\left[\frac{g_1}{m}\right],\nonumber \\
G&=&\frac{3}{16}\ln 3-\frac{1}{2}\ln\left[\frac{g_1}{a}\sqrt{\frac{m}{g_1}}\right].
\end{eqnarray}
\indent For $g_2\neq g_1$, we find two classes of solutions. In the
first class, only $AdS_3\times H^3$ is possible and given by
\begin{eqnarray}
\phi&=&\frac{1}{2}\ln\left[\frac{g_1+g_2}{g_2-g_1}\right],\qquad \sigma=\frac{1}{2}\ln\left[\frac{-3m\sqrt{g_2^2-g_1^2}}{2g_1g_2}\right],\nonumber \\
G&=&\frac{1}{2}\ln\left[-\frac{a(g_2^2-g_1^2)^{\frac{3}{4}}}{g_2}\sqrt{-\frac{3}{2g_1g_2m}}\right]
\end{eqnarray}
where we have chosen $g_1>0$ and $g_2<-g_1$. For positive $g_2$ with
$g_2>g_1$ and $g_1>0$, we find another $AdS_3\times H^3$ solution
\begin{eqnarray}
\phi&=&\frac{1}{2}\ln\left[\frac{g_1+g_2}{g_2-g_1}\right],\qquad \sigma=\frac{1}{2}\ln\left[\frac{3m\sqrt{g_2^2-g_1^2}}{2g_1g_2}\right],\nonumber \\
G&=&\frac{1}{2}\ln\left[\frac{a(g_2^2-g_1^2)^{\frac{3}{4}}}{g_2}\sqrt{-\frac{3}{2g_1g_2m}}\right].
\end{eqnarray}
\indent The second class of solutions is given by
\begin{eqnarray}
\sigma &=&\frac{3}{4}\phi-\frac{1}{4}\ln \left[\frac{(1+e^{6\phi})g_1+(1-e^{6\phi})g_2}{6m}\right],\\
G&=&\phi-\frac{1}{2}\ln\left[\frac{g_2e^{\frac{3\sigma}{2}}(e^{4\phi}-1)}{\lambda a}
\sqrt{\frac{3m}{2[(1+e^{6\phi})g_1+(1-e^{6\phi})g_2]}}\right]
\end{eqnarray}
where $\phi$ is a solution to the following equation
\begin{equation}
(1-3e^{2\phi}-3e^{4\phi}+e^{6\phi})g_1+(1+3e^{2\phi}-3e^{4\phi}-e^{6\phi})g_2=0\, .
\end{equation}
The explicit form of $\phi$ can be written, but we refrain from
giving such a complicated expression. We will however give some
examples of the solutions. Using the relation $g_1=3m$ and choosing
$g_2=\frac{1}{2}g_1$, we find an $AdS_3\times S^3$ solution
characterized by
\begin{equation}
\phi=0.9645,\qquad \sigma=-0.528,\qquad
G=-0.4309-\frac{1}{2}\ln\left[\frac{m}{a}\right].
\end{equation}
It is also possible to obtain an $AdS_3\times H^3$ solution for
$g_2=\frac{1}{2}g_1$. This solution is given by
\begin{equation}
\phi=-0.5732,\qquad \sigma=-0.2723,\qquad
G=-0.5732-\frac{1}{2}\ln\left[\frac{0.3278m}{a}\right].
\end{equation}

\subsection{$N=(1,0)$ two-dimensional SCFTs with $SO(3)\times SO(2)$
symmetry} We now look for a larger residual symmetry. Although the
dilaton $\sigma$ is a singlet of $SO(3)_R\times SO(3)$, the second
$SO(3)$ gauge fields cannot be turned on without turning on some
vector multiplet scalars as can be seen from the supersymmetry
transformation of $\lambda^I_A$. On the other hand, no vector
multiplet scalar is a singlet of the $SO(3)$ symmetry. We can at
most have $SO(3)_R\times SO(2)$ symmetry with $SO(2)$ being a
subgroup of $SO(3)$. Among the $12$ scalars in $SO(4,3)/SO(4)\times
SO(3)$, there is only one $SO(3)_R\times SO(2)$ singlet. This
corresponds to the non-compact generator $Y_{03}$. We then
parametrize the coset representative as
\begin{equation}
L=e^{\Phi Y_{03}}\, .
\end{equation}
The $SO(3)_R$ gauge fields are the same as in the previous case
while the $SO(2)$ gauge field will be chosen to be
\begin{equation}
A^6=b\cos \psi\qquad \textrm{or}\qquad A^6=\frac{b}{y}dx
\end{equation}
for $\Sigma_3=S^3$ and $\Sigma_3=H^3$, respectively.
\\
\indent Apart from the projectors in \eqref{S3_projetor} and
\eqref{H3_projetor}, in this case, we need to impose an additional
projector involving $\gamma_7$ namely
\begin{equation}
\gamma_7\epsilon^A=\sigma_{3\phantom{s}B}^A\epsilon^B\, .
\end{equation}
The critical points would then preserve only half of the
supersymmetry in the previous case. This corresponds to $N=(1,0)$
superconformal symmetry in two dimensions.
\\
\indent With all these and $\lambda=\pm 1$ for $S^3$ and $H^3$,
respectively, we find the following BPS equations
\begin{eqnarray}
\Phi'&=&\frac{1}{2}e^{-3\sigma-\Phi-2G}\left[\lambda b
e^{2\sigma}(1+e^{2\Phi})+2me^{2G}(1-e^{2\Phi})\right],\\
\sigma'&=&\frac{1}{8}e^{-3\sigma-\Phi-2G}\left[6\lambda
e^{2\sigma+\Phi}+\lambda
be^{2\sigma}(e^{2\Phi}-1)\right.\nonumber \\
& &\left.+2e^{2G}(3m-2g_1e^{4\sigma+\Phi}+3me^{2\Phi})\right],\\
G'&=&\frac{1}{8}e^{-3\sigma-\Phi-2G}\left[3\lambda
be^{2\sigma}(e^{2\Phi}-1)+10\lambda
ae^{2\sigma+\Phi}\right.\nonumber \\
& &\left.+2e^{2G}(2g_1e^{4\sigma+\Phi}+m+me^{2\Phi})\right],\\
F'&=&\frac{1}{8}e^{-3\sigma-\Phi-2G}\left[-\lambda
be^{2\sigma}(e^{2\Phi}-1)-6\lambda
ae^{2\sigma+\Phi}\right.\nonumber \\
& &\left.+2e^{2G}(2g_1e^{4\sigma+\Phi}+m+me^{2\Phi})\right].
\end{eqnarray}
When $\Phi=0$ and $b=0$, the above equations reduce to the pure
$F(4)$ gauged supergravity which admits only $AdS_3\times H^3$
solutions in agreement with \cite{F4_nunez}.
\\
\indent We find an $AdS_3\times \Sigma_3$ fixed point given by
\begin{eqnarray}
\sigma&=&\frac{1}{4}\ln
\left[\frac{e^{-\Phi}[3ae^\Phi(e^{2\Phi}-1)+2bm(1+e^{2\Phi}+e^{4\Phi})]}{g_1b(1+e^{2\Phi})}\right],\\
G&=&\frac{1}{4}\ln\left[\frac{be^{-\Phi}(1+e^{2\Phi})[3ae^\Phi(e^{2\Phi}-1)+2b(1+e^{2\Phi}+e^{4\Phi})]}
{4g_1m(e^{2\Phi}-1)^2}\right],\\
\Phi&=&\ln\left[\frac{\sqrt{2a^2+2b^2+2a\sqrt{a^2-2b^2}}-a\pm
\sqrt{a^2-2b^2}}{2b}\right].
\end{eqnarray}
The above solution is written for the $S^3$ case. To find the
solution in the $H^3$ case, $(a,b)$ should be replaced by $(-a,-b)$
in all of the above expressions. The solution is valid for
non-vanishing $a$ and $b$ with $-\sqrt{\frac{a^2}{2}}\leq b\leq
\sqrt{\frac{a^2}{2}}$.
\\
\indent From the analysis, it turns out that only $AdS_3\times H^3$
solutions are possible. As an example of explicit solutions, we take
$a>0$ and choose $b=\pm\frac{a}{\sqrt{2}}$. The solutions are given
by
\begin{eqnarray}
\Phi&=&\ln\left[\frac{\sqrt{3}-1}{\sqrt{2}}\right],\qquad
\sigma=\frac{1}{4}\ln\left[\frac{2\sqrt{2}m(2-\sqrt{3})}{(2\sqrt{3}-3)g_1}\right],\nonumber
\\
G&=&\ln\left[\left(\frac{3}{8}\right)^{\frac{7}{32}}\sqrt{\frac{a}{m}}
\left(\frac{m}{g_1}\right)^{\frac{1}{16}}\right]
\end{eqnarray}
for $b=\frac{a}{\sqrt{2}}$ and
\begin{eqnarray}
\Phi&=&\ln\left[\frac{\sqrt{3}+1}{\sqrt{2}}\right],\qquad
\sigma=\frac{1}{4}\ln\left[\frac{2\sqrt{2}m}{\sqrt{3}g_1}\right],\nonumber
\\
G&=&\ln\left[\left(\frac{3}{8}\right)^{\frac{1}{8}}\sqrt{\frac{a}{m}}
\left(\frac{m}{g_1}\right)^{\frac{1}{4}}\right]
\end{eqnarray}
for $b=-\frac{a}{\sqrt{2}}$. The solution in this case does not have
an analogue in seven dimensions since there is no $SO(3)_R$ scalar
in that case.

\section{Conclusions}\label{conclusion}
We have classified supersymmetric $AdS_4$ and $AdS_3$ solutions of
$N=(1,1)$ six-dimensional gauged supergravity coupled to three
vector multiplets with $SU(2)\times SU(2)$ gauge group. Depending on
the values of the two gauge couplings, there are both $AdS_4\times
S^2$ and $AdS_4\times H^2$ solutions with $SO(2)\times SO(2)$
symmetry and $AdS_4\times H^2$ solutions with $SO(2)$ symmetry. All
solutions preserve eight of the original sixteen supercharges and
are dual to $N=2$ SCFTs in three dimensions. For $AdS_3\times
\Sigma_3$ solutions, we have found new $AdS_3\times S^3$ and
$AdS_3\times H^3$ solutions preserving four supercharges and $SO(3)$
symmetry. These solutions correspond to $N=(1,1)$ SCFTs in two
dimensions. For $SO(3)\times SO(2)$ symmetry, only $AdS_3\times H^3$
solutions exist with $\frac{1}{8}$ supersymmetry unbroken. These
solutions provide gravity duals of $N=(1,0)$ SCFTs in two
dimensions. Apart from the $SO(3)\times SO(2)$ $AdS_3$ fixed point,
the solutions are very similar to those of $N=2$ $SO(4)$ gauged
supergravity in seven dimensions \cite{AdS5_4_N2_7D}.
\\
\indent All of these solutions correspond to IR fixed points of
five-dimensional SCFTs with global symmetry $SU(2)$ in lower
dimensional space-time. There should be RG flows describing twisted
compactifications of these SCFTs on $2$ or $3$-manifolds giving rise
to these $AdS_4$ and $AdS_3$ geometries in the IR. We have not been
able to find analytic solutions for these flows, but numerical
solutions can be obtained as in other cases, see for example
\cite{AdS5_4_N2_7D}. The results obtained in this paper are
hopefully useful in the holographic study of five-dimensional SCFTs
and their compactifications as well as the classification of vacua
of the half-maximal gauged supergravity in six dimensions.
\\
\indent It would be interesting to find a possible embedding of
these solutions in higher dimensions in particular in massive type
IIA supergravity similar to the embedding of pure $F(4)$ gauged
supergravity \cite{Pope_6D_massiveIIA} or in type IIB supergravity
as in \cite{AdS6_Tomasiello} and \cite{Eoin_F4_IIB}. This could give
an interpretation to these solutions in terms of wrapped D4-branes.
However, since there is only one class of known $AdS_6$ solutions,
as shown in \cite{AdS6_from10D}, embedding the $AdS_6$ solutions
with different $SU(2)$ gauge coupling constants (if possible) might
not be straightforward in massive type IIA theory.
\\
\indent It is also interesting to find dual field theories to the
$AdS_4$ and $AdS_3$ critical points identified here. Another
investigation would be to study other types of gauge groups such as
non-compact gauge groups to the matter-coupled $F(4)$ gauged
supergravity and classify all possible gauge groups that admit
supersymmetric $AdS_6$ vacuum similar to the recent analysis in
seven dimensions \cite{Non_compact_7DN2,Jan_N2_7D}. Finally, gravity
solutions with a non-vanishing two-form field could be of interest.
A simple $AdS_3\times R^3$ solution with only the two-form and the
dilaton turned on has been studied in \cite{F4_nunez}. It might be
interesting to study this type of solutions and a more general twist
involving $B_{\mu\nu}$ field within the framework of the matter
coupled gauged supergravity. We leave these issues for future works.
\acknowledgments This work is supported by Chulalongkorn University
through Ratchadapisek Sompoch Endowment Fund under grant Sci-Super
2014-001. The author would like to give a special thank to Block 7
Company Limited where some parts of this work have been done.
\appendix
\section{Supersymmetric $AdS_6$ critical points and holographic RG flows}
In this appendix, we review a description of supersymmetric $AdS_6$
critical points with $SU(2)\times SU(2)$ and $SU(2)_{\textrm{diag}}$
symmetries. We consider only the $SU(2)_{\textrm{diag}}$ singlet
scalar corresponding to the non-compact generator $Y_s$ defined by
\begin{equation}
Y_s=Y_{11}+Y_{22}+Y_{33}\, .
\end{equation}
The $SO(4,3)/SO(4)\times SO(3)$ coset representative is accordingly
parametrized by
\begin{equation}
L=e^{\phi Y_s}\, .\label{L_SO3D}
\end{equation}
The scalar potential can be computed to be
\begin{eqnarray}
V&=&\frac{1}{16}e^{2\sigma}\left[(g_1^2+g_2^2)[\cosh(6\phi)-9\cosh(2\phi)]+8(g_2^2-g_1^2)+8g_1g_2
\sinh^3(2\phi)\right]\nonumber \\
& &+e^{-6\sigma}m^2-4e^{-2\sigma}m(g_1\cosh^3\phi-g_2\sinh^3\phi).
\end{eqnarray}
\indent We are mainly interested in supersymmetric critical points.
Therefore, we set up the BPS equations from supersymmetry
transformations of fermions given in \eqref{delta_psi},
\eqref{delta_chi} and \eqref{delta_lambda} with all but the metric
and scalars $\sigma$ and $\phi$ vanishing. The six-dimensional
matric is taken to be the standard domain wall
\begin{equation}
ds^2=e^{2A(r)}dx^2_{1,4}+dr^2
\end{equation}
where $dx^2_{1,4}$ is the metric on five-dimensional Minkowski
space.
\\
\indent With the projection condition
$\gamma_{\hat{r}}\epsilon_A=\epsilon_A$, the resulting BPS equations
are given by
\begin{eqnarray}
\phi'&=&-\frac{1}{4}e^{\sigma-3\phi}(e^{4\phi}-1)\left[(1+e^{2\phi})g_1+(1-e^{2\phi})g_2\right],\label{6D_eq1}\\
\sigma'&=&-\frac{1}{16}e^{\sigma-3\phi}\left[(1+e^{2\phi})^3g_1+(1-e^{2\phi})^3g_2\right]
+\frac{3}{2}me^{-3\sigma},\label{6D_eq2}\\
A'&=&\frac{1}{16}e^{\sigma-3\phi}\left[(1+e^{2\phi})^3g_1+(1-e^{2\phi})^3g_2\right]
+\frac{1}{2}me^{-3\sigma}\label{6D_eq3}
\end{eqnarray}
where the $r$-derivative is denoted by $'$. From these equations, it
is clearly seen that there are two supersymmetric critical points
namely
\begin{equation}
\phi=0,\qquad \sigma=\frac{1}{4}\ln
\left[\frac{3m}{g_1}\right],\qquad
V_0=-20m^2\left(\frac{g_1}{3m}\right)^{\frac{3}{2}}\label{SO4_AdS6}
\end{equation}
and
\begin{eqnarray}
\phi&=&\frac{1}{2}\ln \left[\frac{g_1+g_2}{g_2-g_1}\right],\qquad
\sigma=\frac{1}{4}\ln
\left[-\frac{3m\sqrt{g_2^2-g_1^2}}{g_1g_2}\right],\nonumber \\
V_0&=&-20m^2\left[-\frac{g_1g_2}{3m\sqrt{g_2^2-g_1^2}}\right]^{\frac{3}{2}}\,
.\label{SO3_AdS6}
\end{eqnarray}
This critical point is valid for $g_2<-g_1$ when $g_1>0$. For $g_1<0$, we need to take $g_2<g_1$. The full scalar mass spectrum at these two critical points can be found in \cite{F4_flow}.
\\
\indent To find an RG flow solution interpolating between these two
critical points, we solve the above BPS equations for $\phi(r)$,
$\sigma(r)$ and $A(r)$. By defining a new radial coordinate
$\tilde{r}$ via $\frac{d\tilde{r}}{dr}=e^{\sigma-3\phi}$, we can
solve equation \eqref{6D_eq1} for $\phi(\tilde{r})$. The solution is
given implicitly by
\begin{eqnarray}
\tilde{r}&=&\frac{4}{g_1+g_2}\phi-\frac{1}{2g_1}\ln(1-e^{2\phi})-\frac{1}{2g_2}\ln(1+e^{2\phi})\nonumber
\\
&
&+\frac{(g_1-g_2)^2}{g_1g_2(g_1+g_2)}\ln\left[(1+e^{2\phi})g_1+(1-e^{2\phi})g_2\right].
\end{eqnarray}
In the above solution, we have omitted an additive integration
constant which can be removed by shifting the coordinate
$\tilde{r}$.
\\
\indent Combining equations \eqref{6D_eq1} and \eqref{6D_eq2}, we
find
\begin{equation}
\frac{d\sigma}{d\phi}=-\frac{e^{3\phi-4\sigma}\left[24m-e^{4\sigma-3\phi}
[(1+e^{2\phi})^3g_1+(1-e^{2\phi})^3g_2]\right]}
{4(e^{4\phi}-1)\left[(1+e^{2\phi})g_1+(1-e^{2\phi})g_2\right]}
\end{equation}
which can be solved to give a solution for $\sigma$
\begin{equation}
\sigma=\frac{1}{4}\ln\left[\frac{e^{-\phi}(6m+C_1(e^{4\phi}-1))}{(1+e^{2\phi})g_1+(1-e^{2\phi})g_2}\right].
\end{equation}
In order to make this solution interpolate between the two critical
points \eqref{SO4_AdS6} and \eqref{SO3_AdS6}, we choose the constant
$C_1$ to be
\begin{equation}
C_1=-\frac{3m(g_1-g_2)^2}{2g_1g_2}
\end{equation}
which gives the solution for $\sigma$
\begin{equation}
\sigma=\frac{1}{4}\ln
\left[\frac{3me^{-\phi}[(1-e^{2\phi})g_1+(1+e^{2\phi})g_2]}{2g_1g_2}\right].
\end{equation}
By the same procedure, we find the solution for $A(r)$ up to an
additive integration constant that can be absorbed by rescaling the
coordinates in $dx^2_{1,4}$
\begin{equation}
A=\frac{1}{4}\phi-\frac{1}{3}\ln
(1-e^{2\phi})-\frac{1}{4}\ln(1+e^{2\phi})+\frac{1}{3}\ln[(1+e^{2\phi})g_1+(1-e^{2\phi})g_2].
\end{equation}
It should be noted that the critical points and the flow solution
have a similar structure to those in seven dimensions
\cite{7D_flow}.


\end{document}